\documentclass[conference]{IEEEtran}
\IEEEoverridecommandlockouts
\usepackage{cite}
\usepackage{amsmath,amssymb,amsfonts}
\usepackage{accents}
\usepackage[linesnumbered,ruled]{algorithm2e}
\usepackage[noend]{algpseudocode}
\usepackage{graphicx}
\usepackage{textcomp}
\usepackage[table,xcdraw]{xcolor}
\usepackage{xcolor}
\usepackage{comment}
\usepackage{multicol,multirow}
\usepackage[T1]{fontenc}
\def\BibTeX{{\rm B\kern-.05em{\sc i\kern-.025em b}\kern-.08em
    T\kern-.1667em\lower.7ex\hbox{E}\kern-.125emX}}
\usepackage{todonotes}
\usepackage{caption}
\newsavebox{\measurebox}
\usepackage{threeparttable}
\usepackage[font=small]{caption}
\usepackage{enumitem}
\usepackage{subfigure}
\usepackage{dblfloatfix}

\usepackage{hyperref}
\hypersetup{
    colorlinks=false
}

\newcommand*{\vsepfbox}[1]{%
  \begingroup
    \sbox0{\fbox{#1}}%
    \setlength{\fboxrule}{0pt}%
    \mbox{\kern-\fboxsep\fbox{\unhbox0}\kern-\fboxsep}%
  \endgroup
}

\begin{document}

\title{Strategic Safety-Critical Attacks Against an Advanced Driver Assistance System
%
}


\author{
\IEEEauthorblockN{{\fontsize{10.5}{12.6}\selectfont Xugui Zhou, Anna Schmedding\IEEEauthorrefmark{2}, Haotian Ren, Lishan Yang\IEEEauthorrefmark{2}, Philip Schowitz\IEEEauthorrefmark{2}, Evgenia Smirni\IEEEauthorrefmark{2}, Homa Alemzadeh}}
\IEEEauthorblockA{
{University of Virginia},
Charlottesville, VA 22904 
\{xz6cz, hr3xw, ha4d\}@virginia.edu}
\IEEEauthorrefmark{2}{William \& Mary},
Williamsburg, VA 23187
\{akschmed, lyang11, philips, esmirni\}@cs.wm.edu
}

\maketitle

\begin{abstract}
A growing number of vehicles are being transformed into semi-autonomous vehicles (Level 2 autonomy) by relying on advanced driver assistance systems (ADAS) to improve the driving experience. However, the increasing complexity and connectivity of ADAS 
expose the vehicles to safety-critical faults and attacks. 
This paper investigates the resilience of a widely-used ADAS against safety-critical attacks that target the control system at opportune times during different driving scenarios and cause accidents. 
Experimental results show that our proposed Context-Aware attacks can achieve an 83.4\% success rate in causing hazards, 99.7\% of which occur without any warnings. These results highlight the intolerance of ADAS to safety-critical attacks and the importance of timely interventions by human drivers or automated recovery mechanisms to prevent accidents.


%
%

\end{abstract}


\section{Introduction}
Over 3.5 million passenger cars worldwide are equipped with level 2 autonomous driving features such as Automated Lane Centering (ALC), Adaptive Cruise Control (ACC), and lane change assistance~\cite{canalys,SAEroadmap}. With level 2 autonomy, the human driver must always be ready to take over the control of the car at any time. Many past studies have shown that unforeseen faults and/or malicious attacks can cause unsafe operation of the autonomous driver assistance systems (ADAS)
with catastrophic consequences~\cite{rubaiyat2018experimental,banerjee2018hands,jha2019ml, jha2020ml, uber_crash, sato2021dirty}. 

There are a variety of vulnerable components within a vehicle that can be targets for attacks, including Electronic Control Units (ECUs), sensors, in-vehicle networks, and V2X (Vehicle-to-Everything) communication~\cite{koscher2010experimental,checkoway2011comprehensive,AVattacksurvey2021,miller2014survey}. Protecting the in-vehicle communication networks is of particular importance because they transmit sensor data, actuator commands, and other safety-critical information among various components. For example, some ADAS (e.g., OpenPilot from Comma.ai~\cite{openpilot}) are integrated with the control system of existing vehicles by tapping into the Controller Area Network (CAN) bus interface through the On-Board Diagnostics (OBD) II port~\cite{mccord2011automotive}. Additionally, critical information is shared through publisher-subscriber messaging systems, such as ROS~\cite{quigley2009ros,aeberhard2015automated} or Cereal~\cite{cereal} which are shown to be vulnerable to a variety of attacks~\cite{Ros2017Secure}. Despite efforts towards protecting these communication channels using techniques such as encryption and intrusion detection~\cite{cho2016fingerprinting,Zhou2020BTMonitorBI}, these protections either cannot detect attacks of a specific type or frequency or are not implemented in most vehicles on the road due to computational costs and the real-time constraints of the vehicle control systems~\cite{miller2013adventures,AVattacksurvey2021,wahl2013iterative,luo2019time}.   

Recent works on autonomous vehicle (AV) safety and security have focused on assessing the impact of hardware faults and physical attacks on the ML accelerators~\cite{li2017understanding} and inputs~\cite{rubaiyat2018experimental, sato2021dirty}, sensor attacks~\cite{petit2014potential,cao2019adversarial}, and attacks targeting the controller~\cite{iqbal2019towards, ding2021mini}. But less attention has been paid to targeted safety-critical attacks on the ADAS output and actuator commands sent over the vulnerable communication channels that might go undetected by the existing safety mechanisms or cannot be acted on by a human driver.

More recently, studies have shown the benefit of contextual information and dynamic AV models ~\cite{rubaiyat2018experimental, jha2019ml, jha2020ml} in designing effective fault injection and attack strategies that result in high hazard coverage. Machine learning (ML) methods such as Bayesian networks~\cite{jha2019ml}, neural networks~\cite{jha2020ml}, and reinforcement learning~\cite{moradi2020exploring} are used to explore the fault parameter space and identify the most salient fault and attack scenarios with adverse impacts on safety. However, such approaches depend on large amounts of data from random fault injection experiments for model training. 

%
%
%

In this paper, we use an orthogonal \textit{model-driven} approach to the above \textit{data-driven} techniques. Instead of focusing on exploring the entirety of the fault parameter space, 
we focus on a systematic characterization of the effect of the values of the parameter space (e.g., start time and duration of faults) {\em in conjunction with} the dynamic state of the vehicle to
identify the most opportune system contexts to launch the attacks.
We propose a Context-Aware safety-critical attack that can find {\em the most critical context} during a driving scenario to activate attacks that \textit{strategically corrupt} the ADAS outputs, with the goal of (1) maximizing the chance of hazards and (2) causing hazards as soon as possible, before being detected/mitigated by the human driver or the ADAS safety mechanisms. We base this approach on the high-level control-theoretic hazard analysis~\cite{leveson2013stpa} and specification of context-dependent safety requirements~\cite{dsn2021zhou,rubaiyat2018experimental} for a typical ADAS, which is applicable to any ADAS with the same functional and safety specifications. 

We assess the resilience of OpenPilot\cite{openpilot}, a widely-used ADAS, against such safety-critical attacks by demonstrating that system hardware and software components can be exploited, and actuator commands can be strategically modified to implement such malicious attacks and cause highly effective targeted hazards and accidents such as collision with other vehicles or road-side objects.

Our study shows that the proposed Context-Aware strategy judiciously selects the most opportune start times and durations for attacks and  
is efficient in exploiting the safety-critical states of ADAS.
We also find that
lane invasions are common and can happen even without injecting faults, that the forward collision warning is not activated at all during attacks, and that the {\em steering angle} is the most vulnerable target. Our experimental results further highlight the importance of driver alertness for timely intervention and hazard prevention and the importance of robust automated safety mechanisms at the latest computational stage, just before execution on actuators.

\section{Preliminaries}

\begin{figure}[t!]
    \centering
    \includegraphics[width=\columnwidth]{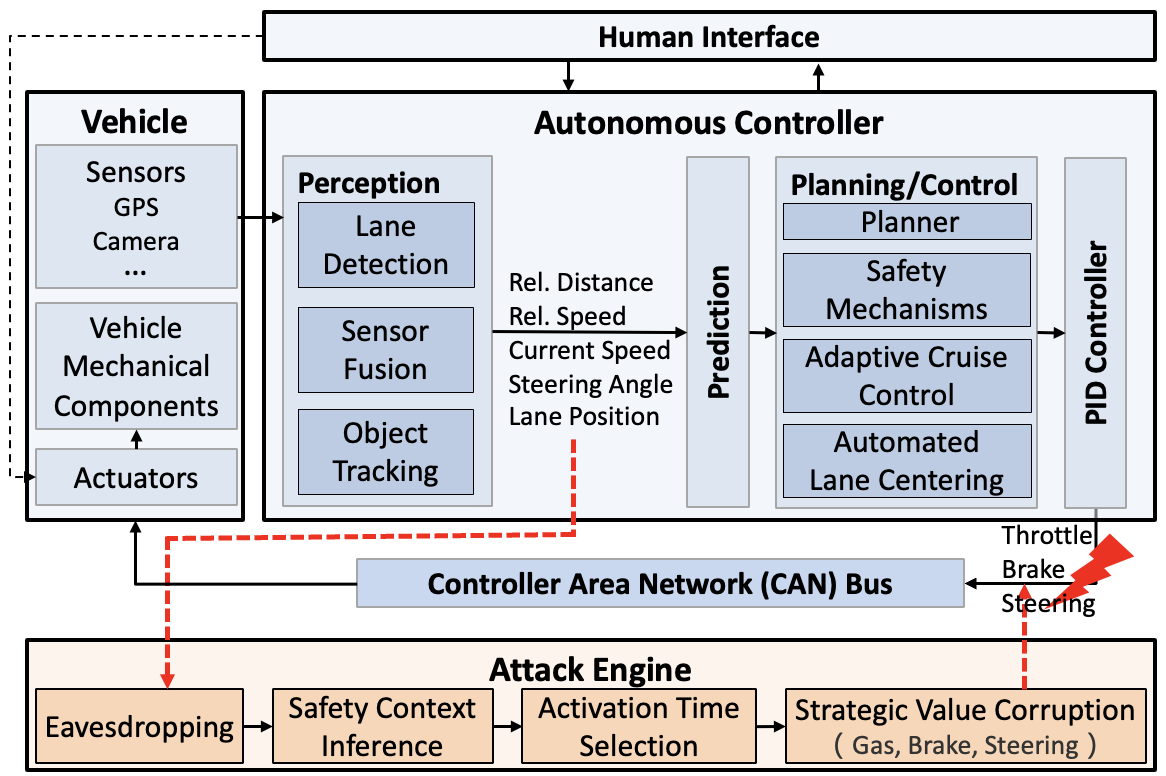}
    \caption{Overview of the control structure of an ADAS with ACC and ALC, and the proposed attack engine. }
    \label{fig:overview}
    \vspace{-1.5em}
\end{figure}

\subsection{Advanced Driver Assistance Systems: OpenPilot}
\label{sec:Openpilot-intro}
Fig.~\ref{fig:overview} presents the overall structure of an ADAS that includes sensing, perception, prediction, planning, and control units. ADAS features such as ALC and ACC rely on machine learning (ML) to perceive the vehicle state and surrounding environment using different sensors (e.g., camera, RADAR, GPS)~\cite{paulo2016ADAS}. Planning and control algorithms are used to adjust gas, brake, and steering to achieve target speed while keeping a safe distance from other vehicles, road lanes, and objects on the road. 

OpenPilot is an alpha quality open-source ADAS that performs the functions of ALC and ACC for over 150 supported car makes and models, including Honda, Toyota, Hyundai, Nissan, Kia, Chrysler, Lexus, Acura, Audi, and VW~\cite{openpilot}. 
In the past six years, over 1,500 monthly active users have collectively driven more than 40 million autonomous miles using OpenPilot. User driving data (e.g., camera, CAN, GPS, operating system logs) are collected by Comma.ai~\cite{comma2k19} and are used to train and test machine learning models.

\textbf{Safety Mechanisms:}
OpenPilot is designed as a fail-safe passive system that requires the driver to be alert at all times. To enforce driver alertness, it also provides a monitoring feature that warns (jolts) the driver when distracted. 

The following safety principles (as required by international standards, e.g., ISO 22179) are incorporated into the design of OpenPilot to ensure that the vehicle does not alter its trajectory too quickly, therefore allowing the driver to safely react~\cite{openpilot_safety}:
\begin{itemize}
    \item The maximum acceleration {\em limit} is set to 2m/s² and maximum deceleration is set to -3.5m/s².
    \item There is a 1-second delay before the vehicle significantly deviates from its original path (e.g., crosses lane lines), allowing the driver time to react to an erroneous steering command. 
    \item The driver can override OpenPilot with minimal effort, i.e., less than 3Nm  
    extra torque on the steering wheel. 
\end{itemize}
OpenPilot and the firmware used in some of the car models controlled by OpenPilot also implement additional automated safety mechanisms such as Forward Collision Warning (FCW)~\cite{openpilot_FCW} and Autonomous Emergency Braking (AEB)~\cite{openpilot_AEB}.

\subsection{Cyber-Physical System Context}
An ADAS is designed by the tight integration of cyber and physical components with a human in the loop. Safety, as an emergent property of Cyber-Physical Systems (CPS), is context-dependent and should be controlled by enforcing constraints on the system behavior and control actions given the overall system state~\cite{leveson2011engineering, dsn2021zhou}. In every control cycle $t$, an ADAS uses sensor measurements to estimate the physical system state $x_{t}$ (e.g., current speed, relative distance to lead vehicle) and decides on a control action, $u_t$, from a finite set of high-level control actions (e.g., \textit{Acceleration}, \textit{Deceleration}, or \textit{Steering}). The high-level control actions issued by ADAS are then translated by a low-level controller into control commands (e.g., \textit{gas} and \textit{brake}) which are sent to the actuators. Upon execution of the control command by the actuators, the physical system transitions to a new state  $x_{t+1}$. 

A sequence of \textit{cyber control actions} $U_{t}=\{u_{t-k+1},...,u_{t-1},$ $u_{t}\}$ issued in $k$ consecutive control cycles is  unsafe if upon its execution in a given state sequence $X_{t}=\{x_{t-k+1},...,x_{t-1},x_{t}\}$, the system  eventually transitions to a state $x_{t'}$ that is hazardous~\cite{dsn2021zhou} (e.g., too close to the lead vehicle). Thus, both the current system state and the control commands issued by  ADAS  contribute to the vehicle safety status. We use this insight to design a Context-Aware safety-critical attack that {\em infers the most critical states} during vehicle operation to \textit{strategically corrupt the control commands} that are sent to the actuators. 

\begin{figure}[t!]
    \centering
    \includegraphics[width=0.7\columnwidth]{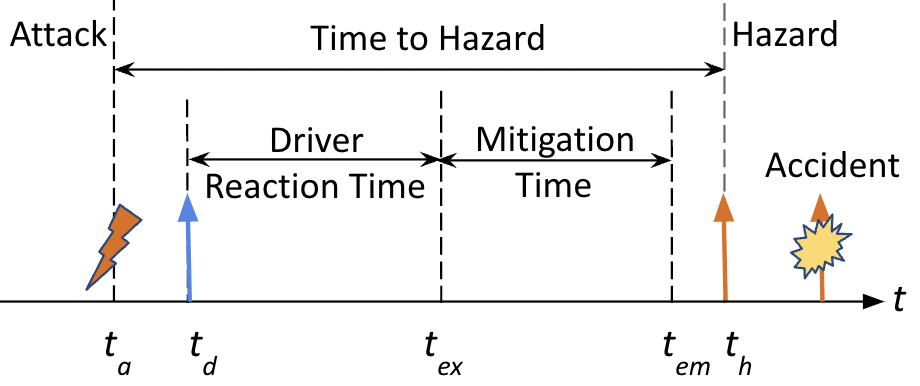}
    \vspace{-0.5em}
    \caption{Timeline of attack propagation. ($t_a$: Attack activated; $t_d$: Attack is detected by the ADAS or the anomaly is sensed by the human driver; $t_{ex}$: Human driver starts to engage; $t_{em}$: End of mitigation; $t_h$: Hazard occurs.)}
    \vspace{-2em}
    \label{fig:timeline}
\end{figure}

Previous studies have shown that there is often a time gap between the activation of faults and 
the final propagation of unsafe control commands to the physical layer, resulting in hazards~\cite{dsn2021zhou,alemzadeh2016targeted}. We define the time between activation of an attack to the occurrence of a hazard as {\textit{Time-to-Hazard (TTH)}} which indicates the maximum time budget for detecting anomalies and engaging in mitigation actions (see Fig.~\ref{fig:timeline}). 

The \textit{Driver Reaction Time} is defined as the time difference between the perception of an alert or anomaly (e.g., seeing an alert raised by the ADAS or recognizing an anomaly) and the start of physically taking an action (e.g., hitting the brake). In the AV literature, the overall driver reaction time (perception and reaction) is reported to be 2.5 seconds on average~\cite{reacttime2019, sato2021dirty}. We define the \textit{Mitigation Time} as the time it takes for any corrective actions (e.g., braking) to be completed. This timing provides a window of opportunity for attackers to cause hazards before being overruled by the human driver or automated safety mechanisms. Fig.~\ref{fig:timeline} shows an example where mitigation successfully completed before the occurrence of the hazard ($t_{em} < t_{h}$). A successful attack should evade detection and/or lead to hazards  before the ADAS or the driver engage ($t_h < t_{ex}$) or complete any mitigation actions (${t_{ex}} {<} t_h<t_{em}$).

\section{Technical Approach}
This section describes the proposed Context-Aware attack strategy and the capabilities and actions needed by an attacker to implement it on OpenPilot.

\vspace{-0.5em}
\subsection{Context-Aware Attack Strategy}
\vspace{-0.5em}

The attacker's goal is to manipulate the control commands, including gas, brake, and steering angle, to maximize the chance of hazard occurrence (e.g., violating longitudinal or lateral safety distance) while avoiding detection by the  ADAS safety mechanisms and the human driver. 
The attacker is interested in causing one of the following accidents:
\begin{itemize}
    \item \textbf{A1:} Collision with the lead vehicle.
    \item \textbf{A2:} Rear-end collision, causing traffic congestion.
    \item \textbf{A3:} Collision with road-side objects or other vehicles in the neighboring lane.
\end{itemize}
by forcing the system to transition into one of the following hazardous states: 
\begin{itemize}
    \item \textbf{H1:} AV violates safe following-distance constraints with the lead vehicle, which may result in A1. 
    \item \textbf{H2:} AV decelerates to a complete stop although there is no lead vehicle, which may lead to A2. 
    \item \textbf{H3:} AV drives out of lane, which may lead to A3.
\end{itemize}

\begin{table}[b!]
\vspace{-1.5em}
\caption{Safety context table for an ADAS with ALC and ACC }
\vspace{-0.5em}
\label{tab:contexttable}
\resizebox{\columnwidth}{!}{%
\begin{threeparttable}
\begin{tabular}{|c|c|c|c|}
\hline
\textbf{Rule} & \textbf{System Context}         & \textbf{Control Action} & \textbf{Potential Hazard} \\ \hline
1    & $HWT \leqslant t_{safe} \wedge RS > 0$ &  $u_1$ & H1 \\ \hline
2    & $HWT > t_{safe} \wedge RS \leqslant 0 \wedge Speed > \beta_1 $ &  $u_2$ & H2 \\ \hline
3    & $d_{left} \leqslant 0.1m \wedge Speed >\beta_2 $ &  $u_3$ & H3 \\ \hline
4    & $d_{right} \leqslant 0.1m \wedge Speed >\beta_2 $ &  $u_4$ & H3 \\ \hline

\end{tabular}

\begin{tablenotes}\footnotesize
\item [*] HWT: Headway Time = Relative Distance/Current Speed; 
\item [*] RS: Relative Speed = Current Speed - Lead Speed; 
\item [*] $d_{left},d_{right}$: Distance to the left/right edge of current lane;
\item[*] $u_{1,2,3,4}$: Acceleration, Deceleration, Steering Left, Steering Right.
\item[*] $t_{safe} \in [2,3]s$, $\beta_1,\beta_2 \in [20,35] mph$

\end{tablenotes}

\end{threeparttable}

}
\end{table}

To increase the chance of hazards, we adopt a control-theoretic hazard analysis method \cite{leveson2013stpa} to identify the specific combinations of system states and control actions that most likely lead to hazards. 
Table~\ref{tab:contexttable} shows an example context table that describes unsafe system contexts, including the specific high-level system context under which specific types of control actions may be unsafe and lead to safety hazards. For example, row 1 states that when the Headway Time is less than a safety limit $t_{safe}$ (e.g., 2s), and the AV speed is faster than that of the lead vehicle ($RS>0$), an $Acceleration$ control action is unsafe as it will result in a forward collision with the lead vehicle. This high-level identification of context-dependent unsafe control actions can be done by an attacker based on the knowledge of typical functionality of an ADAS and be applied to any ADAS with the same functional specification. The unknown thresholds $t_{safe}$, $\beta_1$ and $\beta_2$ can be specified based on domain knowledge or past data~\cite{dsn2021zhou}. 

Our proposed Context-Aware attack strategy uses the critical system contexts described in Table~\ref{tab:contexttable} as the trigger for injecting unsafe control commands~\cite{Av2001attack}.
To evade detection, the control actions generated by the attack must be within the limits that are not noticeable to a human operator and are checked by the ADAS safety mechanisms, while minimizing the Time to Hazard (TTH) (see Fig.~\ref{fig:timeline}) and maximizing the chance of resulting in any hazards. To achieve these goals, the following optimization problem is formulated: 
\vspace{-0.5em}
\begin{align}
\label{eq:opt}
\tiny
    minimize&_{_{TTH}}max\{Pr\{x_{t_{+TTH}} \in Hazardous\}\}\\
    %
    %
    s.t.\ & brake \geq limit_{brake}\nonumber\\
    &accel \leqslant limit_{accel}\nonumber\\
    &\Delta steering < limit_{steer} \nonumber\\
    &\hat{v}_{t+1} \leqslant 1.1v_{cruise} \nonumber \\
    & \hat{v}_{t+1|t} = \hat{v}_{t} +accel*\Delta t 
    \label{eq:dynamic}\\
    & \hat{v}_{t+1} = \hat{v}_{t+1|t} + K_t*(v_{t+1}-\hat{v}_{t+1|t})
    \label{eq:kalman}
\end{align}
where $brake$, $accel$, and $steering$ indicate the modified values of control commands, and $\hat{v}_{t+1}$ represents the predicted speed of the Ego vehicle at the next time step, which can be estimated using Eq. \ref{eq:dynamic} that approximates the dynamics of the vehicle by assuming linear acceleration for a short time period $\Delta t$ (10 ms). 
A Kalman filter~\cite{bishop2001kalman} (with the Kalman Gain parameter $K_t$, see Eq. \ref{eq:kalman}) is used to update the estimation using the measured speed $v_{t+1}$ at the next time step. 
$limit_{accel}$, $limit_{brake}$, $limit_{steer}$ are the constraints on the output control commands defined by the safety checking rules of the target vehicle, including those of its ADAS.



\subsection{Attack Model}
To implement the Context-Aware attack strategy, the attacker needs 1) access to the sensor measurements and/or information shared through the in-vehicle communication network to estimate the current system state and 2) the capability of modifying the actuator commands with faulty values. Possible entry points for executing such malicious actions include 
the wireless networks~\cite{nie2017free}, in-vehicle networks
(e.g., CAN, FlexRay, Ethernet, Bluetooth, or telematics devices) \cite{AVattacksurvey2021,Payne2019CarHA}, the passive keyless entry and start system \cite{Francillon2011keyless}, the vehicle to everything communication, and/or vulnerable components supplied by different vendors \cite{parul2014VANETs,hackBWM2019cai}. 
The attackers can gather information about the system configuration by monitoring and decoding the communication traffic and can identify potential vulnerabilities through publicly available documents such as open-source code \cite{openpilot}.

For example, the attack can be designed based on offline code/data analysis to infer the safety constraints and parameters described in Equations (1)-(3). It can then be implemented as a malware which is deployed via compromising over-the-air updates and, thus, can be activated on multiple cars running the same ADAS to maximize damage. Once deployed, the malware injects malicious commands at critical times by strategically selecting unsafe values to maximize the chance of evading detection by ADAS safety mechanisms while avoiding mitigation by the driver, i.e., hazards occur within a period that is shorter than the driver reaction time.

\subsection{Attack Procedure}
The overall procedure and steps for executing Context-Aware attacks are summarized as follows:

\begin{figure}[t!]
    \centering
    \includegraphics[width=0.95\columnwidth]{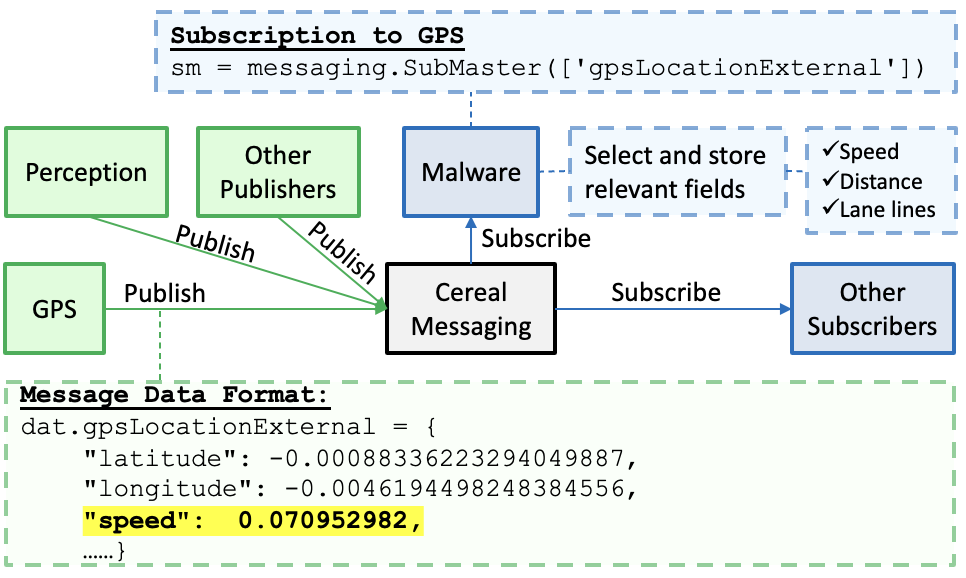}
    \vspace{-0.5em}
    \caption{Cereal messaging eavesdropping.}
    \label{fig:eavesdropping}
    \vspace{-1.5em}
\end{figure}
\textbf{Eavesdropping:}
This step is accomplished by listening to the sensor sockets and the in-vehicle communication network, decoding the messages passed among different software components, and extracting the sensor data and critical state information. 
In OpenPilot, this can be achieved through local or remote subscriptions to the messaging system used for internal packet communication, called Cereal~\cite{cereal}. Cereal is a publisher-subscriber messaging specification for robotic systems (similar to ROS\cite{quigley2009ros}), which is used for publishing messages by sensing and perception modules (e.g., GPS, Radar) and can be subscribed to by other OpenPilot modules (e.g., ACC, ALC) and any malicious software (see Fig.~\ref{fig:eavesdropping}).

Since  OpenPilot is open-source, the format of cereal messages is publicly available \cite{cerealformat}.
An example of eavesdropping on the GPS messages is shown in Fig.~\ref{fig:eavesdropping}.
To extract the information needed for safety context inference, the attacker needs to subscribe to the following events:
    1) ``{\em gpsLocationExternal}'' events to learn the speed of the Ego vehicle published by GPS;
    2) ``{\em modelV2}'' events to receive messages from the perception module to learn the lane line positions; 
    3) ``{\em radarState}'' events published by the RADAR to learn the relative speed and distance of the lead vehicle.

\textbf{Safety Context Inference:} 
\label{sec:contextinfernde}
Next, the attacker uses the basic state information ${x_t}$, including the speed of the Ego vehicle, the vehicle's lateral position, the lane line positions, and the relative distance to the lead vehicle, to infer the more complex and human-interpretable state variables described in the safety specification (Table~\ref{tab:contexttable}). 
For example, the headway time (HWT) is an important metric for identifying critical system context and can be calculated based on the Ego vehicle's current speed and relative distance to the lead vehicle. 

\textbf{Attack Type and Activation Time Selection:}
A context matcher detects whether the current system state matches any of the critical system contexts specified in Table~\ref{tab:contexttable}. If a matching case is found, the attack engine decides on the attack action (e.g., {\em Acceleration}) based on the unsafe action specified for the context and activates the attack. Table~\ref{tab:FIintro} lists the attack types to be activated based on different contexts. If two different context conditions are simultaneously detected, both control actions (e.g., {\em Acceleration} and {\em Steering}) are activated.

\textbf{Strategic Value Corruption:} 
In the last step, the selected attack type (e.g., {\em Acceleration}) is translated into low-level control commands (e.g., maximum {\em gas} and zero {\em brake}). The attack engine dynamically corrupts the control command values to not exceed the safety limits checked by OpenPilot's safety mechanisms (see Eq. 1-3). These safety limits are identified and encoded offline based on open-source code and publicly available documentations. Table~\ref{tab:FI-strategies} shows the specific safety limits we used for the attack types shown in Table~\ref{tab:FIintro}. 
\begin{figure}[t]
    \centering
    \includegraphics[width=.95\columnwidth]{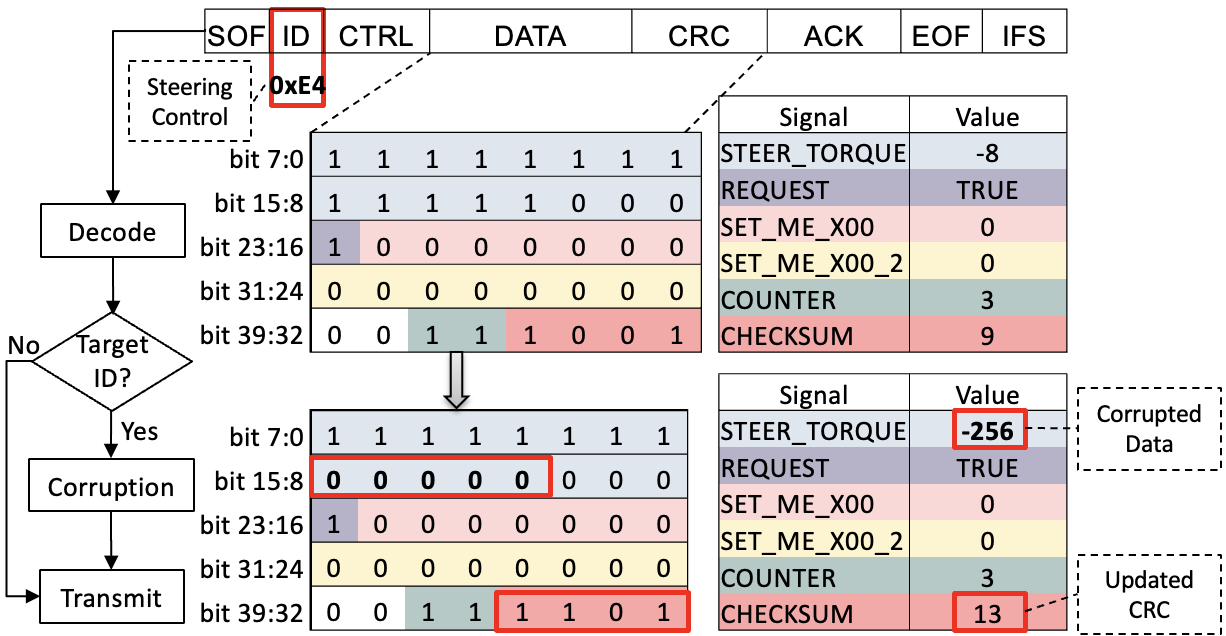}
    \vspace{-0.5em}
    \caption{An example of changing a steering output CAN message.}
     \vspace{-1.5em}
    \label{fig:encodeCAN}
\end{figure}

Finally, the faulty commands are sent to the target actuators by manipulating CAN messages. 
The information in a CAN bus message can be decoded using reverse engineering and the open-source Database Container (DBC)~\cite{autopi-dbc-explained} configuration~\cite{commaai-opendbc} of a specific car model. 
The attack engine then corrupts the specific CAN message that carries a target control command using the command's unique identifier (e.g., 0xE4 for steering as shown in Fig.~\ref{fig:encodeCAN}). 
The attacker also updates the checksum after corrupting targeted control commands, so the integrity of the corrupted CAN message is maintained.

\section{Experiments}
\begin{figure}[t!]
    \centering
    \includegraphics[width=0.9\columnwidth]{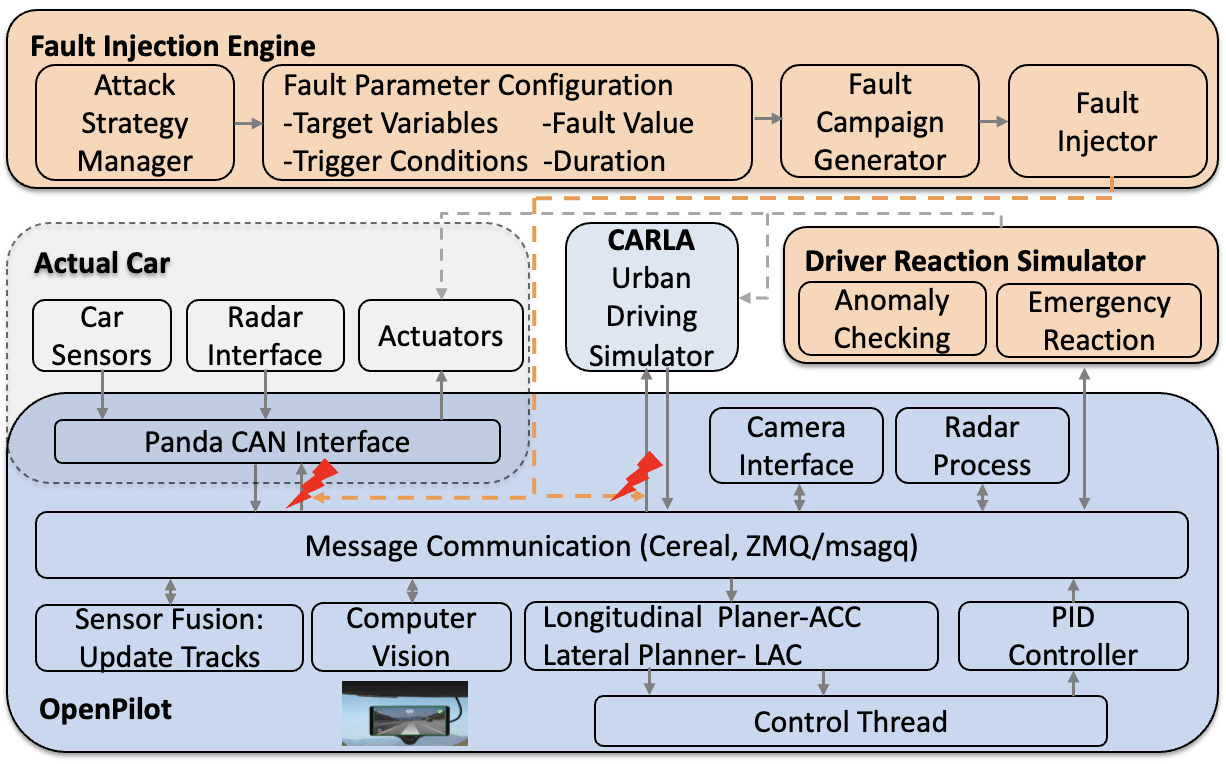}
    \vspace{-0.5em}
    \caption{Overall architecture of OpenPilot, integrated with CARLA, the driver reaction simulator, and the fault injection engine. [Online Available: \url{https://github.com/UVA-DSA/openpilot-CARLA}].}
    \label{fig:OPCARLA}
    \vspace{-1.5em}
\end{figure}

To evaluate the effectiveness of the proposed Context-Aware attack, we develop a simulation platform consisting of the OpenPilot control software integrated with the CARLA urban driving simulator~\cite{Dosovitskiy17}, a driver reaction simulator, and a software-implemented fault injection engine. The platform architecture is shown in Fig.~\ref{fig:OPCARLA} and is described next. 

The OpenPilot safety mechanisms (see Section \ref{sec:Openpilot-intro}) are implemented in its control software and the Panda CAN interface. Panda is a universal OBD adapter developed by Comma.ai~\cite{Panda} that provides access to almost all car sensors through the CAN bus. When integrated with the CARLA driving simulator, the Panda software and hardware are not utilized by OpenPilot. Therefore, Panda safety checks are not enforced. Here, we consider all safety limits checked by Panda as constraints for generating faulty values for the Context-Aware attacks so that they evade detection by Panda when it is engaged in actual driving (Eq. \ref{eq:opt}).

Our experiments are done on Ubuntu 20.04 LTS, with OpenPilot v0.8.9 and CARLA v9.11. A single simulation of OpenPilot contains 5000 time-steps, each step lasts about 10 ms, which in total equals 50 seconds. 

\subsection{Driving Scenarios}
Using the CARLA simulator, we create different driving scenarios where the Ego vehicle, cruising at 60mph from 50, 70, or 100 meters away, approaches a lead vehicle with different behaviors as follows.

\begin{itemize}
    \item S1: Lead vehicle cruises at the speed of 35 mph;
    \item S2: Lead vehicle cruises at the speed of 50 mph;
    \item S3: Lead vehicle slows down from an initial speed of 50 mph to 35 mph;
    \item S4: Lead vehicle accelerates from an initial speed of 35 mph to 50 mph.
\end{itemize}

\begin{figure}[b]
    \vspace{-1.5em}
    \centering
	\begin{minipage}{0.5\columnwidth}
	\scriptsize \centering
    \includegraphics[width=\columnwidth]{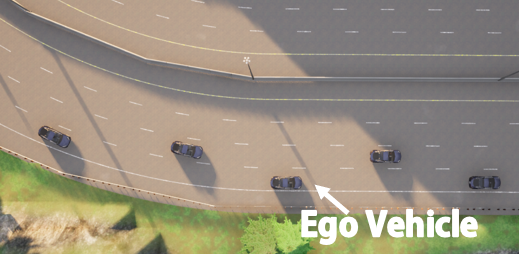}
    
    (a)  An example initial position of Ego Vehicle (EV) and other reference vehicles.
    \end{minipage}
    \hfill
    	\begin{minipage}{0.45\columnwidth}\scriptsize \centering
    \includegraphics[width=\columnwidth]{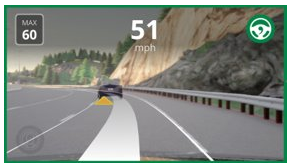}
    
    (b)   The user interface of OpenPilot during the simulation.
    \end{minipage}
    
    	\begin{minipage}{0.51\columnwidth}\scriptsize \centering
    \includegraphics[width=\columnwidth]{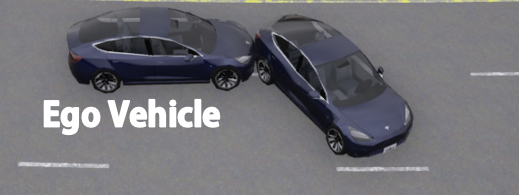}
    
    (c)  EV collides with the lead vehicle.
    \end{minipage}\hfill
    	\begin{minipage}{0.45\columnwidth}\scriptsize \centering
    \includegraphics[width=\columnwidth]{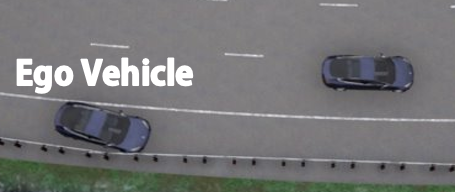}
    
    (d) EV collides with the guardrail.
    \end{minipage}
    \caption{  Driving scenarios in OpenPilot.}
    \label{fig:opUI}
\end{figure}

Fig.~\ref{fig:opUI}(a-b) show different views of a simulated scenario. 

\subsection{Driver Reaction Simulator}
To mimic the situation where the human driver takes over the control of the vehicle in an emergency situation, we designed a driver reaction simulator (see Fig. \ref{fig:OPCARLA}). 
The simulated driver is alerted when the ADAS raises any safety alarms (e.g., FCW) or the driver observes any anomalies in the vehicle status that last for a noticeable period of time (i.e., at least 1 second). Anomalies include hard brake ($|Brake|>|limit_{brake}|$), unexpected increase in acceleration ($Accel>limit_{accel}$) or steering ($Steering>limit_{steer}$), or the vehicle speed exceeding cruising speed by more than 10\% ($Speed>1.1v_{cruise}$). To make the attack more challenging, anomalies that occur within one simulation time step (10ms) attract the driver's attention.
The driver physically takes action after 2.5 seconds (the average driver reaction time) to process an alert or anomaly~\cite{reacttime2019}).

Typically human drivers respond to sudden unintended acceleration with a hard brake within 1.5 seconds. We model this using an exponential function that approximates the general brake curve as follows \cite{Driverbrake2019}:
\begin{equation}
\label{eq:brakesim}
    brake = e^{10t-12}/(1+e^{10t-12})
\end{equation}
We also apply the same reaction to sudden steering. 
The attack engine stops the attack as soon as the driver engages.  



\subsection{Attack Types}
For each driving scenario, we simulate six types of attacks (see Table~\ref{tab:FIintro}) by injecting faults into each output variable as well as their combinations. For example, for the \textit{Acceleration-Steering} attack, we inject faults to \textit{Gas} and \textit{Steering Angle} (either left or right angle). We limit the injected values within the acceptable ranges by the OpenPilot control software. 
Each scenario is tested with three different initial  positions for the lead vehicle and is repeated 20 times
to capture variations due to changes in the simulated driving environment and attack timings.
This results in 60 (20 $\times$ 3) simulations per attack type and a total number of 1,440 (60 $\times$ 6 $\times$ 4) for all simulated attacks and driving scenarios.
\begin{table}[t]
\caption{Fault injection experiments}
\vspace{-0.5em}
\label{tab:FIintro}
\resizebox{\columnwidth}{!}{%
\begin{tabular}{|l|c|c|c|c|}
\hline
\textbf{Attack Type} & \textbf{Accel} & \textbf{Brake} & \textbf{Steering Angle} & \textbf{No. Attacks} \\ \hline
Acceleration & $limit_{accel}$ & 0 & \cellcolor[HTML]{BFBFBF} & 60 \\ \hline
Deceleration & 0 & $limit_{brake}$ & \cellcolor[HTML]{BFBFBF} & 60 \\ \hline
Steering-Left & \cellcolor[HTML]{BFBFBF} & \cellcolor[HTML]{BFBFBF} & -$limit_{steer}$ & 60 \\ \hline
Steering-Right & \cellcolor[HTML]{BFBFBF} & \cellcolor[HTML]{BFBFBF} & $limit_{steer}$ & 60 \\ \hline
Acceleration-Steering & $limit_{accel}$ & 0 & ±$limit_{steer}$ & 60 \\ \hline
Deceleration-Steering & 0 & $limit_{brake}$ & ±$limit_{steer}$ & 60 \\ \hline
\end{tabular}%



}
\vspace{-2em}

\end{table}


    
    

\begin{table}[b]
\vspace{-1.5em}

\caption{Overview of attack strategies}
\vspace{-0.5em}
\label{tab:FI-strategies}
\resizebox{\columnwidth}{!}{%
\begin{threeparttable}
\begin{tabular}{|l|l|l|l|c|}
\hline
\textbf{Attack Strategy} & \textbf{Start Time} & \textbf{Duration} & \textbf{Attack Values} & \textbf{No. Attacks} \\ \hline
Random-ST+DUR & Uniform [5,40]s & Uniform [0.5,2.5]s & {Fixed\textsuperscript{1}} & {14,400} \\ \hline
Random-ST & Uniform [5,40]s & 2.5s & Fixed & 1,440 \\ \hline
Random-DUR & Context-Aware & Uniform [0.5,2.5]s & Fixed & 1,440 \\ \hline
Context-Aware & Context-Aware & Context-Aware & Strategic$\textsuperscript{2}$ & 1,440 \\ \hline
\end{tabular}%

\begin{tablenotes}\footnotesize
\item [1] Fixed: use the maximum limit of each output command defined in OpenPilot: $limit_{steer}=0.5^{\circ}$, $limit_{brake}=-4m/s^2$, $limit_{accel}=2.4 m/s^2$.
\item [2] Strategic: dynamically choose the attack value according to Eq. \ref{eq:opt}-\ref{eq:kalman} ($limit_{steer}=0.25^{\circ}$, $limit_{brake}=-3.5m/s^2$, $limit_{accel}=2 m/s^2$).
\end{tablenotes}

\end{threeparttable}
}
\end{table}

\subsection{Baselines}
In addition to the Context-Aware strategy, we design three baseline strategies to test the ADAS resilience to different attacks (see Table~\ref{tab:FI-strategies}). 
The first baseline (referred to as \textit{Random-ST+DUR}) uses a random start time uniformly distributed within [5, 40] seconds (5 seconds after the start of simulation till 10 seconds before the end),
with attack duration uniformly distributed within [0.5, 2.5] seconds. We run \textit{Random-ST+DUR} strategy for 14,400 simulations to maximize coverage of the critical attack parameters.
For the second baseline (referred to as \textit{Random-ST}), we randomly choose a start time but fix the attack duration to be equal to the average driver reaction time (2.5 seconds). To test the relationship between hazards and attack duration, we also design a third baseline (referred to as \textit{Random-DUR}) by randomly choosing the attack duration from a range of [0.5, 2.5] seconds with the start time inferred based on context. 
All the attack values are within the range of OpenPilot safety checks. Note that aggressive random attacks (e.g., bombarding the CAN-bus with out-of-the-range values) may get detected by existing intrusion detection mechanisms for in-vehicular networks \cite{cho2016fingerprinting,AVattacksurvey2021} and the OpenPilot safety checks, so they are not considered here.

\subsection{Results}

\subsubsection{System Resilience Evaluation}
\begin{figure}[b]
\vspace{-2.1em}
    \centering
    	\begin{minipage}{0.85\columnwidth}
	\scriptsize \centering
    \includegraphics[width=\columnwidth]{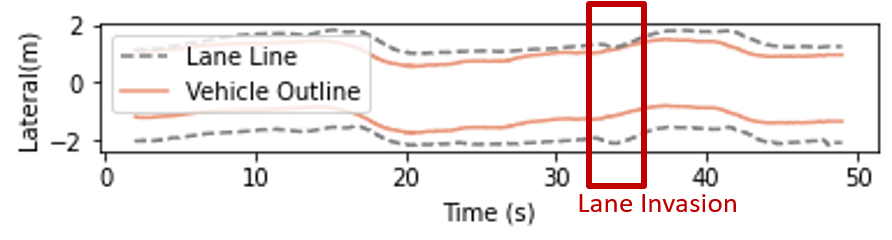}
    \end{minipage}
    
    
    \vspace{-0.5em}
    \caption{Trajectory of the Ego Vehicle during an attack-free simulation.} 
    \label{fig:ALC}
\end{figure}

We evaluate the resilience of OpenPilot in presence of an alert driver by running the simulations with and without the attacks. 
Table~\ref{tab:totalresult} shows that under normal system operation, when no attacks are engaged, no hazards or accidents occur. However, 2 \textit{steer saturated} alerts were raised due to the steering angle exceeding the pre-defined safety limits in OpenPilot.
Fig.~\ref{fig:ALC} shows an example of the performance of the ALC system.
We observe that the ALC system does not keep the Ego vehicle in the center of the lane at all times, and lane invasions occur with an average frequency of 0.46 times per second, which can lead to out-of-lane hazards or collision with road-side objects. This indicates that the ALC and ACC systems do not cooperate well, which is a defect in the control software and needs to be fixed.

\noindent\vsepfbox{
    \parbox{0.95\linewidth}{
        \textbf{Observation 1: Lane invasions can happen even without any attacks.} 
    }
}

\begin{table}[t]
\caption{Attack strategy comparisons with an alert driver.}
\vspace{-0.5em}
\label{tab:totalresult}
\resizebox{\columnwidth}{!}{%
\begin{tabular}{|l|c|c|c|c|c|c|}
\hline
\textbf{Attack Strategy} & \textbf{\begin{tabular}[c]{@{}c@{}}Alerts\end{tabular}} & \textbf{\begin{tabular}[c]{@{}c@{}}Hazards\end{tabular}} & \textbf{\begin{tabular}[c]{@{}c@{}}Accident\end{tabular}} & \textbf{\begin{tabular}[c]{@{}c@{}}Hazards\& \\ no Alerts\end{tabular}} & \multicolumn{1}{l|}{\textbf{\begin{tabular}[c]{@{}l@{}}LaneInvasion\\ (No. Event/s)\end{tabular}}}& \textbf{\begin{tabular}[c]{@{}c@{}}TTH(s) \\ (Avg. ± Std.)\end{tabular}} \\ \hline
\textbf{No Attacks} & \begin{tabular}[c]{@{}c@{}}2\\ (0.1\%)\end{tabular} & 0 & 0 & 0 & 0.46&\cellcolor[HTML]{BFBFBF}  \\ \hline
\textbf{Random-ST+DUR} & \begin{tabular}[c]{@{}c@{}}3248 \\ (22.6\%)\end{tabular} & \begin{tabular}[c]{@{}c@{}}5727\\ (39.8\%)\end{tabular} & \begin{tabular}[c]{@{}c@{}} 3293\\ (22.9\%)\end{tabular} & \begin{tabular}[c]{@{}c@{}} 3083 \\ (21.4\%)\end{tabular} & 1.03 & 1.61±1.96\\ \hline




\textbf{Random-ST}     & \begin{tabular}[c]{@{}c@{}}346\\ (24.0\%)\end{tabular}          & \begin{tabular}[c]{@{}c@{}}771\\ (53.5\%)\end{tabular}          & \begin{tabular}[c]{@{}c@{}}516\\ (35.8\%)\end{tabular}           & \begin{tabular}[c]{@{}c@{}}474\\ (32.9\%)\end{tabular}                  & 0.68                                                                         & {1.49±0.73    }        \\ \hline
\textbf{Random\_DUR}   & {\begin{tabular}[c]{@{}c@{}}210\\ (14.6\%)\end{tabular}  }        &{ \begin{tabular}[c]{@{}c@{}}388\\ (26.9\%)\end{tabular}  }        &{ \begin{tabular}[c]{@{}c@{}}332\\ (23.1\%)\end{tabular}  }         &{ \begin{tabular}[c]{@{}c@{}}229\\ (15.9\%)\end{tabular}  }                &{ 0.46                                                 }                         &{ 1.92±1.17    }        \\ \hline

\textbf{Context-Aware} & \begin{tabular}[c]{@{}c@{}}4\\ (0.3\%)\end{tabular} & \begin{tabular}[c]{@{}c@{}}1201\\ (83.4\%)\end{tabular} & \begin{tabular}[c]{@{}c@{}}641\\ (44.5\%)\end{tabular} & \begin{tabular}[c]{@{}c@{}}1197\\ (83.1\%)\end{tabular} & 0.66 & 2.43±1.29\\ \hline

\end{tabular}%
}
\vspace{-2.0em}
\end{table}

\subsubsection{Comparison to Random Attack Strategies}
Table \ref{tab:totalresult} also shows that the Context-Aware strategy outperforms the three random strategies and achieves the highest hazard coverage of 
{83.4\%, with 99.7\% (1197/1201)}
of hazards occurring without any alerts. Note that 
{53.4\% (641/1201)} of hazards result in accidents, 
including collision with the lead vehicle and road-side objects (see Fig.~\ref{fig:opUI}(c-d)). In these cases, the alert raised by ADAS is the \textit{steer saturated} warning, while the more relevant {\em forward collision} warning (FCW) is not activated as the brake output is kept less than the safety threshold of  OpenPilot. 
We also observe an increased number of lane invasions per second for almost all attacks due to the occurrence of out-of-lane hazards. Despite achieving the highest hazard coverage,  the Context-Aware attacks keep the number of lane invasions and alerts low because of the strategic value corruption.

\noindent\vsepfbox{
    \parbox{0.95\linewidth}{
        \textbf{Observation 2: The Context-Aware attack strategy is efficient in exploiting safety critical states of ADAS. During attacks, the {\em forward collision} warning does not get activated at all.} 
    }
}

From Table \ref{tab:totalresult} we also see that the average TTH of Context-Aware attack is larger than the Random attacks due to a higher hazard rate in \textit{Acceleration} attack that has a longer TTH. 

\begin{figure}[b]
        \vspace{-2.1em}

     \centering
        \includegraphics[width=0.8\columnwidth]{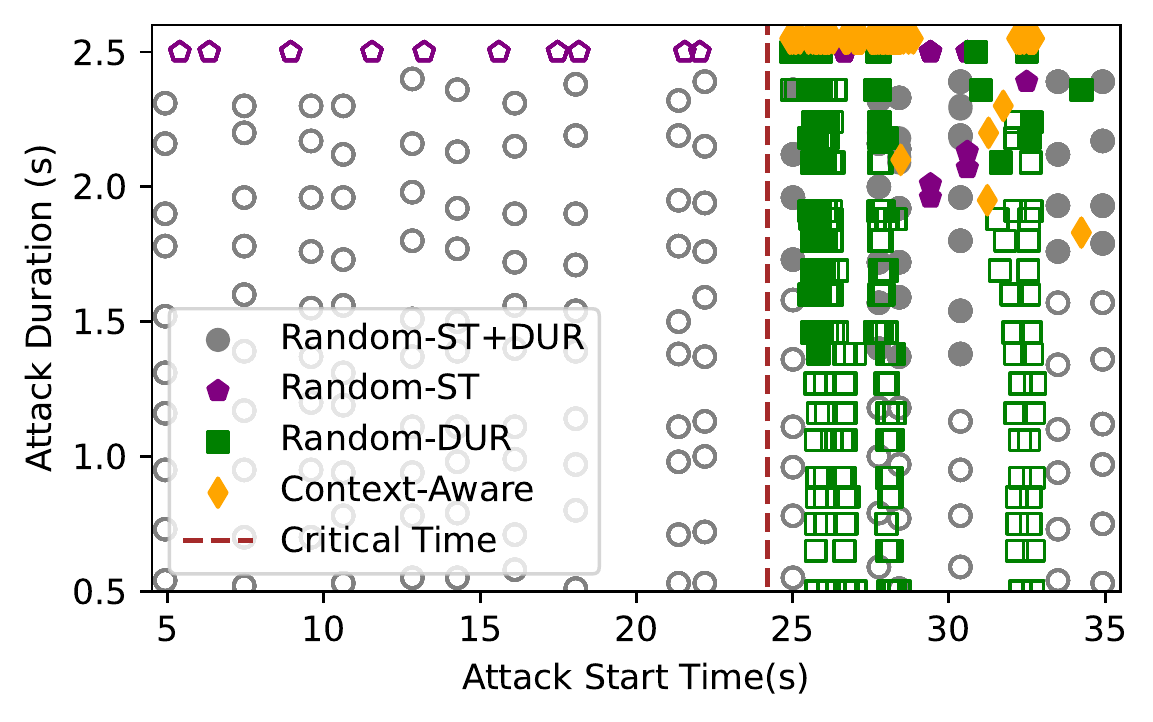}

        \vspace{-0.5em}
        \caption{State space of ``Attack start time" and ``Duration" for {\em Acceleration} attacks (solid shapes correspond to hazardous results and empty ones to non-hazardous).}

        \label{fig:duration}
\end{figure}

\begin{table*}[t]
\caption{Context-aware attack with or without strategic value corruption and with an alert driver.}
\vspace{-0.5em}
\label{tab:context-newlimits}
\label{tab:summarywithfaulttypes}
\resizebox{\textwidth}{!}{%
\begin{threeparttable}

\begin{tabular}{|c|cccccccc||ccccc|}
\hline
\multirow{3}{*}{\textbf{Attack Type}} & \multicolumn{8}{c||}{\textbf{No Strategic Value Corruption}} & \multicolumn{5}{c|}{\textbf{With Strategic Value Corruption}} \\ \cline{2-14} 
 & \multicolumn{1}{c|}{\multirow{2}{*}{\textbf{Alerts}}} & \multicolumn{1}{c|}{\multirow{2}{*}{\textbf{Hazards}}} & \multicolumn{1}{c|}{\multirow{2}{*}{\textbf{Accident}}} & \multicolumn{1}{c|}{\multirow{2}{*}{\textbf{\begin{tabular}[c]{@{}c@{}}TTH(s) \\ (Avg. ± Std.)\end{tabular}}}} & \multicolumn{1}{c|}{\multirow{2}{*}{\textbf{\begin{tabular}[c]{@{}c@{}}Prevented\\ Hazards\end{tabular}}}} & \multicolumn{1}{c|}{\multirow{2}{*}{\textbf{\begin{tabular}[c]{@{}c@{}}New \\ Hazards\end{tabular}}}} & \multicolumn{1}{c|}{\multirow{2}{*}{\textbf{\begin{tabular}[c]{@{}c@{}}Prevented\\ Accidents\end{tabular}}}} & \multirow{2}{*}{\textbf{\begin{tabular}[c]{@{}c@{}}Reduced \\ Accidents\end{tabular}}} & \multicolumn{1}{c|}{\multirow{2}{*}{\textbf{Alerts}}} & \multicolumn{1}{c|}{\multirow{2}{*}{\textbf{Hazards}}} & \multicolumn{1}{c|}{\multirow{2}{*}{\textbf{Accident}}} & \multicolumn{1}{c|}{\multirow{2}{*}{\textbf{\begin{tabular}[c]{@{}c@{}}TTH(s) \\ (Avg. ± Std.)\end{tabular}}}} & \multirow{2}{*}{\textbf{\begin{tabular}[c]{@{}c@{}}Driver \\ Prevention*\end{tabular}}} \\
 & \multicolumn{1}{c|}{} & \multicolumn{1}{c|}{} & \multicolumn{1}{c|}{} & \multicolumn{1}{c|}{} & \multicolumn{1}{c|}{} & \multicolumn{1}{c|}{} & \multicolumn{1}{c|}{} &  & \multicolumn{1}{c|}{} & \multicolumn{1}{c|}{} & \multicolumn{1}{c|}{} & \multicolumn{1}{c|}{} &  \\ \hline
Acceleration & \multicolumn{1}{c|}{\begin{tabular}[c]{@{}c@{}}4 \\ (1.7\%)\end{tabular}} & \multicolumn{1}{c|}{\begin{tabular}[c]{@{}c@{}}200\\ (83.3\%)\end{tabular}} & \multicolumn{1}{c|}{\begin{tabular}[c]{@{}c@{}}120\\ (50.0\%)\end{tabular}} & \multicolumn{1}{c|}{3.33±0.23} & \multicolumn{1}{c|}{\begin{tabular}[c]{@{}c@{}}200\\ (83.3\%)\end{tabular}} & \multicolumn{1}{c|}{\begin{tabular}[c]{@{}c@{}}160\\ (66.7\%)\end{tabular}} & \multicolumn{1}{c|}{\begin{tabular}[c]{@{}c@{}}200\\ (83.3\%)\end{tabular}} & \begin{tabular}[c]{@{}c@{}}120\\ (50\%)\end{tabular} & \multicolumn{1}{c|}{\begin{tabular}[c]{@{}c@{}}1\\ (0.4\%)\end{tabular}} & \multicolumn{1}{c|}{\begin{tabular}[c]{@{}c@{}}160\\ (66.7\%)\end{tabular}} & \multicolumn{1}{c|}{\begin{tabular}[c]{@{}c@{}}160\\ (66.7\%)\end{tabular}} & \multicolumn{1}{c|}{5.03±1.22} & 1/1 \\ \hline
Deceleration & \multicolumn{1}{c|}{\begin{tabular}[c]{@{}c@{}}1\\ (0.4\%)\end{tabular}} & \multicolumn{1}{c|}{\begin{tabular}[c]{@{}c@{}}99 \\ (41.2\%)\end{tabular}} & \multicolumn{1}{c|}{\begin{tabular}[c]{@{}c@{}}0\\ (0.0\%)\end{tabular}} & \multicolumn{1}{c|}{2.62±0.04} & \multicolumn{1}{c|}{\begin{tabular}[c]{@{}c@{}}141\\ (58.8\%)\end{tabular}} & \multicolumn{1}{c|}{0} & \multicolumn{1}{c|}{0} & 0 & \multicolumn{1}{c|}{\begin{tabular}[c]{@{}c@{}}0\\ (0.0\%)\end{tabular}} & \multicolumn{1}{c|}{\begin{tabular}[c]{@{}c@{}}231\\ (96.2\%)\end{tabular}} & \multicolumn{1}{c|}{\begin{tabular}[c]{@{}c@{}}0\\ (0.0\%)\end{tabular}} & \multicolumn{1}{c|}{2.77±0.10} & 0/0 \\ \hline
Steering-Left & \multicolumn{1}{c|}{\begin{tabular}[c]{@{}c@{}}122\\ (50.8\%)\end{tabular}} & \multicolumn{1}{c|}{\begin{tabular}[c]{@{}c@{}}187\\ (77.9\%)\end{tabular}} & \multicolumn{1}{c|}{\begin{tabular}[c]{@{}c@{}}175\\ (72.9\%)\end{tabular}} & \multicolumn{1}{c|}{1.11±0.86} & \multicolumn{1}{c|}{0} & \multicolumn{1}{c|}{0} & \multicolumn{1}{c|}{0} & 0 & \multicolumn{1}{c|}{\begin{tabular}[c]{@{}c@{}}1\\ (0.4\%)\end{tabular}} & \multicolumn{1}{c|}{\begin{tabular}[c]{@{}c@{}}90\\ (37.5\%)\end{tabular}} & \multicolumn{1}{c|}{\begin{tabular}[c]{@{}c@{}}1\\ (0.4\%)\end{tabular}} & \multicolumn{1}{c|}{1.33±0.17} & 0/0 \\ \hline
Steering-Right & \multicolumn{1}{c|}{\begin{tabular}[c]{@{}c@{}}2\\ (0.8\%)\end{tabular}} & \multicolumn{1}{c|}{\begin{tabular}[c]{@{}c@{}}240\\ (100.0\%)\end{tabular}} & \multicolumn{1}{c|}{\begin{tabular}[c]{@{}c@{}}240\\ (100.0\%)\end{tabular}} & \multicolumn{1}{c|}{1.63±0.08} & \multicolumn{1}{c|}{0} & \multicolumn{1}{c|}{0} & \multicolumn{1}{c|}{0} & 0 & \multicolumn{1}{c|}{\begin{tabular}[c]{@{}c@{}}0\\ (0.0\%)\end{tabular}} & \multicolumn{1}{c|}{\begin{tabular}[c]{@{}c@{}}240\\ (100.0\%)\end{tabular}} & \multicolumn{1}{c|}{\begin{tabular}[c]{@{}c@{}}240\\ (100.0\%)\end{tabular}} & \multicolumn{1}{c|}{1.39±0.10} & 0/0 \\ \hline
Acceleration-Steering & \multicolumn{1}{c|}{\begin{tabular}[c]{@{}c@{}}2\\ (0.8\%)\end{tabular}} & \multicolumn{1}{c|}{\begin{tabular}[c]{@{}c@{}}240\\ (100.0\%)\end{tabular}} & \multicolumn{1}{c|}{\begin{tabular}[c]{@{}c@{}}240\\ (100.0\%)\end{tabular}} & \multicolumn{1}{c|}{1.51±0.15} & \multicolumn{1}{c|}{0} & \multicolumn{1}{c|}{0} & \multicolumn{1}{c|}{0} & \begin{tabular}[c]{@{}c@{}}-1\\ (0.4\%)\end{tabular} & \multicolumn{1}{c|}{\begin{tabular}[c]{@{}c@{}}2\\ (0.8\%)\end{tabular}} & \multicolumn{1}{c|}{\begin{tabular}[c]{@{}c@{}}240\\ (100.0\%)\end{tabular}} & \multicolumn{1}{c|}{\begin{tabular}[c]{@{}c@{}}240\\ (100.0\%)\end{tabular}} & \multicolumn{1}{c|}{1.47±0.26} & 0/0 \\ \hline
Deceleration-Steering & \multicolumn{1}{c|}{\begin{tabular}[c]{@{}c@{}}3\\ (1.2\%)\end{tabular}} & \multicolumn{1}{c|}{\begin{tabular}[c]{@{}c@{}}138\\ (57.5\%)\end{tabular}} & \multicolumn{1}{c|}{\begin{tabular}[c]{@{}c@{}}17\\ (7.1\%)\end{tabular}} & \multicolumn{1}{c|}{2.63±0.02} & \multicolumn{1}{c|}{\begin{tabular}[c]{@{}c@{}}170\\ (70.8\%)\end{tabular}} & \multicolumn{1}{c|}{\begin{tabular}[c]{@{}c@{}}68\\ (28.3\%)\end{tabular}} & \multicolumn{1}{c|}{0} & \begin{tabular}[c]{@{}c@{}}-17\\ (7.1\%)\end{tabular} & \multicolumn{1}{c|}{\begin{tabular}[c]{@{}c@{}}0\\ (0.0\%)\end{tabular}} & \multicolumn{1}{c|}{\begin{tabular}[c]{@{}c@{}}240\\ (100.0\%)\end{tabular}} & \multicolumn{1}{c|}{\begin{tabular}[c]{@{}c@{}}0\\ (0.0\%)\end{tabular}} & \multicolumn{1}{c|}{2.77±0.06} & 0/0 \\ \hline
Total & \multicolumn{1}{c|}{\begin{tabular}[c]{@{}c@{}}142\\ (9.9\%)\end{tabular}} & \multicolumn{1}{c|}{\begin{tabular}[c]{@{}c@{}}1104\\ (76.6\%)\end{tabular}} & \multicolumn{1}{c|}{\begin{tabular}[c]{@{}c@{}}792\\ (55.0\%)\end{tabular}} & \multicolumn{1}{c|}{2.04±1.10} & \multicolumn{1}{c|}{\begin{tabular}[c]{@{}c@{}}511\\ (36.8\%)\end{tabular}} & \multicolumn{1}{c|}{\begin{tabular}[c]{@{}c@{}}228\\ (16.4\%)\end{tabular}} & \multicolumn{1}{c|}{\begin{tabular}[c]{@{}c@{}}200\\ (22.4\%)\end{tabular}} & \begin{tabular}[c]{@{}c@{}}102\\ (11.4\%)\end{tabular} & \multicolumn{1}{c|}{\begin{tabular}[c]{@{}c@{}}4\\ (0.3\%)\end{tabular}} & \multicolumn{1}{c|}{\begin{tabular}[c]{@{}c@{}}1201\\ (83.4\%)\end{tabular}} & \multicolumn{1}{c|}{\begin{tabular}[c]{@{}c@{}}641\\ (44.5\%)\end{tabular}} & \multicolumn{1}{c|}{2.43±1.29} & 1/1 \\ \hline
\end{tabular}%

\begin{tablenotes}\footnotesize
\item [*] The number of hazards/accidents prevented when a human driver simulator is added in the simulation. 

\end{tablenotes}

\end{threeparttable}

}
\vspace{-2em}
\end{table*}

\subsubsection{Evaluation of Attack Duration and Start Time}

To further evaluate the importance of attack {\em duration} and {\em start} time, 
we assess the coverage of the fault parameter space by different attack strategies. 
Fig.~\ref{fig:duration} illustrates a sample parameter space for durations between 0.5 to 2.5 seconds and start times between 5 to 35 seconds for the \textit{Acceleration} attack type. Each dot in this figure represents an attack simulation. The solid dots represent simulations with hazards. This figure illustrates that an attack does not result in any hazard if not activated within a critical time window (after dashed line at about 24-25 seconds), regardless of how long the attack lasts. After finding the critical launch moment, the attack needs to last for a time period (at least 1.5 seconds) to cause a hazard.  Therefore, it is important to find both the opportune time to start an attack and the required duration to increase the hazard success rate.

We also observe that the Context-Aware strategy (marked by orange diamonds) are all solid (hazardous) and located within the critical time window. The dots that correspond to Random-ST and Random-DUR strategies result in a significant number of non-hazardous cases.
This figure further attests the efficiency of the proposed Context-Aware strategy. 


\noindent\vsepfbox{
    \parbox{0.95\linewidth}{
        \textbf{Observation 3: Context-Aware selection of start time and duration does not waste resources on non-hazardous random injections.} 
    }
}

\subsubsection{Evaluation of the Strategic Value Corruption} 
In this set of experiments, we further evaluate the performance of the Context-Aware strategy with and without the strategic value selection. 
Note that the attacks without strategic value corruption may be detected by Panda's safety checks, if deployed on an actual vehicle. But Panda's safety checks could also be disabled by the attacker or bypassed if the attack is launched after the safety checks (e.g., on the OBD II port). 

Table~\ref{tab:summarywithfaulttypes} shows the results across different attack types, including the number of hazards that are prevented by the driver. For timely hazard mitigation, the driver reaction/mitigation times should be shorter than the Time-to-Hazard (TTH) (See average TTHs in Table \ref{tab:summarywithfaulttypes}). Our experiments indicate that without the driver reaction, the attacks without strategic value corruption could achieve very high hazard and accident success rates (almost 100\% for all attack types; not shown in the table due to space limits). 
However, when simulating the human driver reaction, 83.3\% of hazards are prevented for the \textit{Acceleration} attack, reducing 50\% of collision events. Similar hazard reductions are observed for the \textit{Deceleration} (58.8\%) and \textit{Deceleration-Steering} (70.8\%) attacks.

\noindent\vsepfbox{
    \parbox{0.95\linewidth}{
        \textbf{Observation 4: Human alertness for timely intervention is important in preventing hazards and accidents.} 
    }
}

However, the driver reaction does not prevent {\em Steering} attacks (zero hazards prevented for steering attacks in Table~\ref{tab:summarywithfaulttypes}), and these attacks still achieve very high hazard and accident success rates (e.g., 100\% for \textit{Steering-Right} and \textit{Acceleration-Steering}). This is because hazards happen in less than 1.63s, which is much less than the average human driver reaction time (2.5s), indicating that attacks targeting the steering angle are the most difficult to be mitigated by the driver. 
It should be noted that the \textit{Steering-Left} attacks achieve lower success in causing hazards compared to \textit{Steering-Right} attacks (77.9\% vs. 100\%) because the Ego vehicle is initialized to a lane closer to the right guardrail while it travels on a left-curved road. 


\noindent\vsepfbox{
    \parbox{0.95\linewidth}{
        \textbf{Observation 5: Steering is the most effective attack type that cannot be easily halted by the human driver.} 
    }
}

Although driver intervention reduces hazard and accident rates, it may also introduce new hazards. For example, to avoid a collision with the lead vehicle, the Ego vehicle may stop in the middle of a lane causing a rear collision, or collide with curb objects. 
Table~\ref{tab:summarywithfaulttypes} shows that up to 66.7\% new hazards happened after preventing attacks on the gas output. 

After adding the strategic value corruption, even though there is an overall 6.8\% increase in hazard success rates (76.6\% to 83.4\%), the total number of alerts generated by the ADAS decreases to 4, and only less than 0.1\% of the induced hazards are prevented by the driver, even for the cases where the average TTH is longer than the average driver reaction time (2.5s) (e.g., \textit{Acceleration}, \textit{Deceleration} and \textit{Deceleration-Steering} attacks). This further illustrates the effectiveness of the Context-Aware strategy for evading detection by the ADAS and/or the human driver.



\noindent\vsepfbox{
    \parbox{0.95\linewidth}{
        \textbf{Observation 6: The strategic value corruption is effective in evading human driver detection and safety checks of ADAS.} 
    }
}

\section{Threats to Validity}


Although our attack strategy is efficient in finding potential weaknesses in the ADAS control software, its robustness and efficacy might be affected by the quality of sensor data used for context inference or by existing defense mechanisms (e.g., control invariant detection~\cite{choi2018detecting} or context-aware monitoring \cite{dsn2021zhou}).
Our simulations consider OpenPilot safety checks and human driver interventions. But other safety and security mechanisms that can be implemented in firmware or hardware interface of the car (e.g., Panda's safety checks, AEB, encryption, or intrusion detection) are not included in this study.
Further evaluation of the robustness and detectability of the attacks are directions of future work. 

\section{Conclusion}

This paper presents a strategic Context-Aware attack that targets control commands within an ADAS. This attack finds the most critical times during a driving scenario to activate attacks, as well as the attack durations, which can cause hazards before a human driver or the ADAS safety mechanism can correct the behavior. The proposed attack is efficient since large numbers of random fault injections are not needed to guide the approach.
Our experimental results and observations show that steering is particularly vulnerable and that the existing warning system for forward collisions is insufficient. Our results also highlight the importance of human alertness for timely intervention
for preventing hazards and accidents, and the importance of
automated safety mechanisms that can check the control actions
issued by ADAS.

\section*{Acknowledgment}
 This work was partially supported by the Commonwealth of Virginia under Grant CoVA CCI: C-Q122-WM-02 and by the National Science Foundation (NSF) under Grant No. 1748737.

\bibliographystyle{IEEEtran}
\bibliography{main}

\end{document}